# Mirror actively deformed and regulated for applications in space: design and performance

Marie Laslandes
Emmanuel Hugot
Marc Ferrari
Claire Hourtoule
Christian Singer
Christophe Devilliers
Céline Lopez
Frédéric Chazallet

# Mirror actively deformed and regulated for applications in space: design and performance

**Marie Laslandes**
**Emmanuel Hugot**
**Marc Ferrari**
**Claire Hourtoule**
Aix Marseille Université
CNRS, LAM (Laboratoire d'Astrophysique de Marseille)
UMR 7326, 13388
Marseille, France
E-mail: marie.laslandes@oamp.fr

**Christian Singer**
**Christophe Devilliers**
Thales Alenia Space
100 Boulevard du Midi, BP.99
06156 Cannes la Bocca, France

**Céline Lopez**
Thales SESO
305 rue Louis Armand
13290 Aix en Provence, France

**Frédéric Chazallet**
Shakti
27 Boulevard Charles Moretti
13014 Marseille, France

**Abstract.** The need for both high quality images and lightweight structures is one of the main drivers in space telescope design. An efficient wavefront control system will become mandatory in future large observatories, retaining performance while relaxing specifications in the global system's stability. We present the mirror actively deformed and regulated for applications in space project, which aims to demonstrate the applicability of active optics for future space instrumentation. It has led to the development of a 24-actuator, 90-mm-diameter active mirror, able to compensate for large lightweight primary mirror deformations in the telescope's exit pupil. The correcting system has been designed for expected wavefront errors from 3-m-class lightweight primary mirrors, while also taking into account constraints for space use. Finite element analysis allowed an optimization of the system in order to achieve a precision of correction better than 10 nm rms. A dedicated testbed has been designed to fully characterize the integrated system performance in representative operating conditions. It is composed of: a telescope simulator, an active correction loop, a point spread function imager, and a Fizeau interferometer. All conducted tests demonstrated the correcting mirror performance and has improved this technology maturity to a TRL4. © *The Authors. Published by SPIE under a Creative Commons Attribution 3.0 Unported License. Distribution or reproduction of this work in whole or in part requires full attribution of the original publication, including its DOI.*
[DOI: 10.1117/1.OE.52.9.091803]

## 1 Introduction

Advancements in space telescope technologies have allowed for significant breakthroughs in our understanding of astrophysical and terrestrial phenomena. The instrument requirements depend on the science objectives. For instance, exoplanet direct imaging and characterization requires a combination of high spatial resolution, excellent sensitivity, high contrast, and extreme stability.[1,2] In the field of cosmology, dark matter characterization requires large surveys, with a wide field of view and extremely precise astrometry.[3] Earth observations require a large field of view, high angular resolution, wide spectral range, and eventually rapid revisit capability.[4] Thus, the common needs that drive telescope evolution are higher angular resolution, sensitivity and wavefront stability.

As angular resolution and sensitivity depend directly on the optical aperture,[5] larger primary mirrors must be used. One of the key aspects for space telescope technology is then to increase the diameter while keeping mass and size suitable for launch, as well as structural stability and optical surface quality in-flight.[6,7]

The most famous large space observatory is the Hubble Space Telescope.[8] Launched in 1990, it operates in the ultraviolet/visible. Its primary mirror is 2.4 m in diameter and is $F/24$. It has been polished to a surface error of 6.3 nm rms. Made of ULE glass with an areal density of 180 kg/m$^2$, this mirror presents exceptional stiffness and thermal stability. The largest telescope currently flying is the Herschel Space Observatory, in orbit since 2009.[9,10] Its primary mirror is 3.5 m in diameter and is $F/0.5$. These dimensions have been chosen to be as large as possible with launch fairing constraints. The mirror is made of SiC with an areal density of 21.8 kg/m$^2$. The telescope operates in the infrared, between 55 and 672 $\mu$m, with an optical prescription for the surface error of 3 $\mu$m rms. Larger monolithic mirrors will not fit in the actual launch fairing, so segmented telescope concepts must be adopted. The James Webb Space Telescope, planned to be launched in 2018, will be a major step forward for space telescopes.[11] Its primary mirror is 6.5 m in diameter and is $F/1.2$. The telescope operates in the infrared, between 0.6 and 2.7 $\mu$m, and it will be passively cooled down to 40 K thanks to its large sunshield. The large sizes of the primary mirror and sunshield, as well as the distance between the primary and secondary mirrors, require the observatory to be folded up in the launch fairing and deployed in flight. The primary mirror is composed of 18 beryllium segments mounted on an active structure which will be unfolded. The areal density of the mirror assembly is 50 kg/m$^2$. JWST is designed to be diffraction-limited at 2 $\mu$m, so the surface error requirement for each segment is 25 nm rms. To meet this requirement, the mirrors will be actively controlled through seven degrees of freedom per segment.[12]

These three examples showcase the state-of-the-art for space telescopes. As future space telescopes evolve towards larger diameters and more compact architectures, without a significant mass increase, they will be more sensitive to thermo-elastic, gravitational effects and misalignment, making the achievement of an ultrastable structure difficult.[13,14] Starting with JWST, the next generation of space telescopes will require an active mirror in order to meet surface quality and stability requirements. The larger the telescope and the smaller the wavelength, the more important stability constraints will be, which requires the development of innovative active systems.

For about 30 years, developments in active and adaptive optics allow an efficient wavefront error (WFE) correction in large ground-based observatories, in order to reach the telescope diffraction limit.[15] On the one hand, adaptive optics systems analyze atmospheric turbulence effects and correct them with one or more deformable mirrors (DM).[16] On the other hand, active optics compensate the large mirrors' thermo-elastic and gravity deformations: the optimal shape is maintained with push/pull actuators located behind the optical surface.[17,18] Moreover, the growing complexity of optical instrumentation requires innovative systems. In this context, active optics are useful for variable optical path instruments where an active mirror will allow the correction, *in situ* and in real time, of aberrations evolving with the instrument's configuration.[19,20] Dedicated to the correction of optical aberrations induced by the instrument's intrinsic defects, active optics techniques could be efficiently employed in future spaceborne telescopes. However, existing active systems are not directly applicable for space instrumentation; considerations about weight, size, power consumption, mechanical strength and reliability must first be addressed.

So, active optics will allow a technological breakthrough by providing means to ensure optical quality in the future large telescopes.[21] First, it will correct a constant bias linked to the difference of gravity between integration on Earth at 1 g and operation in space at 0 g, but also average thermal environment and alignment errors. Secondly, it will compensate for thermo-elastic deformation of the telescope structure and primary mirror, due to thermal fluctuations linked to the orbital dynamics. In this context, two different approaches are being studied to compensate for large, lightweight mirrors' deformation in space. The first solution consists in maintaining the primary mirror's optimal shape with numerous actuators under the optical surface. This approach has been adopted for the JWST, and several studies are developing active lightweight mirrors where numerous actuators are embedded in the substrate.[22,23] Such an active primary mirror would allow for a reconfigurable telescope architecture: segments can be moved and their shapes adapted depending on their positions.[24] The second solution consists of performing the correction in the exit pupil plane of the telescope, further along the optical train. It requires a small and light active mirror with a limited number of actuators. A prototype mirror with 24 actuators, designed for this second approach, is presented in this paper. It has been developed in the framework of the mirror actively deformed and regulated for applications in space (MADRAS) project. The system has been optimized with finite element analysis (FEA) and its performance has been characterized in a laboratory environment, improving its Technology Readiness Level to TRL4.[25]

## 2 Active Correcting Mirror Conception

### 2.1 Specification of the MADRAS Project

Starting from existing telescope deformation data, a system study has been performed in order to model the expected deformation maps of 3-m-class primary mirrors under external perturbations such as gravity and temperature changes. They are defined according to the application (Earth observation in low or geostationary orbit or astronomical observation at high angular resolution), and according to the telescope type (monolithic or deployable). A synthesis of these different needs has allowed the definition of a common specification for the MADRAS mirror.

The error budget provides four main items which limit image quality: mirror manufacturing; the assembly, integration, and testing; gravity effects; and on-orbit thermo-elastic deformation. The expected deformations are decomposed into Zernike polynomials, giving the number of Zernike modes to be corrected, along with their amplitudes and required precisions. The mirror's manufacturing will induce high spatial frequency errors which are not addressed by the active correcting system. Integration and alignment errors will mostly induce third order aberrations. Gravity and thermo-elastic effects will deform the mirror in the low Zernike modes, up to the fifth order. The specifications for the MADRAS system are defined by including all significant Zernike terms from these four error sources, presented in Table 1. Considering that rigid body motions will be corrected by a separate system, for example a five degrees of freedom gimbal on the secondary mirror; tip, tilt and focus modes are not addressed here. In order to meet the required image quality in the visible, the residual WFE after correction must be less than 5 nm rms for each modes separately and less than 10 nm rms for a global WFE, composed of a combination of these modes.

The correction will be performed in an optical plane conjugated to the primary mirror, so that the active mirror is 90 mm diameter, which is the size of the exit pupil plane of the considered Korsch designs.[26] The weight of the correcting system is limited to 5 kg. The system's reliability and robustness are studied, and the overall system is designed to survive both space and launch environments. However, the validation of the actuators and the electronics is a separate study, so the MADRAS project does not address this. For the performance demonstration, wavefront control is performed through Shack–Hartmann wavefront sensing, but this strategy could be adapted depending on the application. As the effects of zero gravity and thermal dilatation are addressed, the required actuation frequency is low, the demonstrator operates at 1 Hz.

### 2.2 Optimization of Mirror Geometry

Designed to generate the specified modes, the MADRAS mirror has 24 actuators. The system has been optimized and fully characterized with FEA before manufacturing and integration. The main drivers for the design were the precision of correction, the weight, and the mechanical strength, notably during launch.

#### 2.2.1 Multimode deformable mirror design

The chosen mirror geometry has been developed by Lemaître.[27] As presented in Fig. 1, it is a monolithic piece of Zerodur, made of a circular pupil with an external thicker ring and 12 arms. This design is based on the similarity between the Zernike polynomials used in optical aberration theory[28] and the Clebsch polynomials used in elasticity theory.[29] The application of 24 discrete forces on both extremities of each arm allows the generation of Zernikes defined by $n = m$ and $n = m + 2$; $n$ and $m$ being the radial and azimuthal polynomials' orders. With 12 arms, $m$ is included between 0 and 6. In addition a central clamp holds the system, notably for launch, and allows the generation of spherical aberration ($m = 0$ and $n = 4$).

**Table 1** MADRAS correction specification and corrective performance of each specified mode: residual wavefront error (WFE) deduced from finite element analysis (FEA), from interferometric measurements and measured in closed-loop and maximum stress in Zerodur.

| Mode | Incoming WFE (nm rms) | Residual WFE | | | Stress max (from FEA) (MPa) |
|---|---|---|---|---|---|
| | | FEA (nm rms) | Interferometry (nm rms) | Closed-loop (nm rms) | |
| Coma3 | 200 | 0.7 | 1.8 | 6.2 | 1.02 |
| Astigmatism3 | 150 | 2.5 | 2.5 | 3.3 | 0.14 |
| Spherical3 | 50 | 5.8 | 5.8 | 7.9 | 0.25 |
| Trefoil5 | 30 | 0.2 | 0.3 | 1.0 | 0.10 |
| Astigmatism5 | 30 | 1.8 | 2.2 | 2.9 | 0.36 |
| Tetrafoil7 | 30 | 1.4 | 1.6 | 1.1 | 0.27 |
| Trefoil7 | 30 | 1.6 | 2.2 | 3.3 | 0.77 |
| Pentafoil9 | 30 | 6.8 | 6.9 | 7.3 | 0.61 |
| Tetrafoil9 | 30 | 4.5 | 4.5 | 4.8 | 1.58 |
| Worst case | 265.3 | 10.7 | 11.1 | 14.6 | <5.10 |

In this design, the forces that deform the mirror are applied far from the optical surface. This way, it avoids actuator print-through.[30] Moreover, it decouples the number of actuators from the mirror diameter: the number of required actuators is only driven by the maximal spatial frequency to be corrected. Finally, this design is not limited to a single actuator technology: any actuator applying discrete forces can be used with this type of mirror.

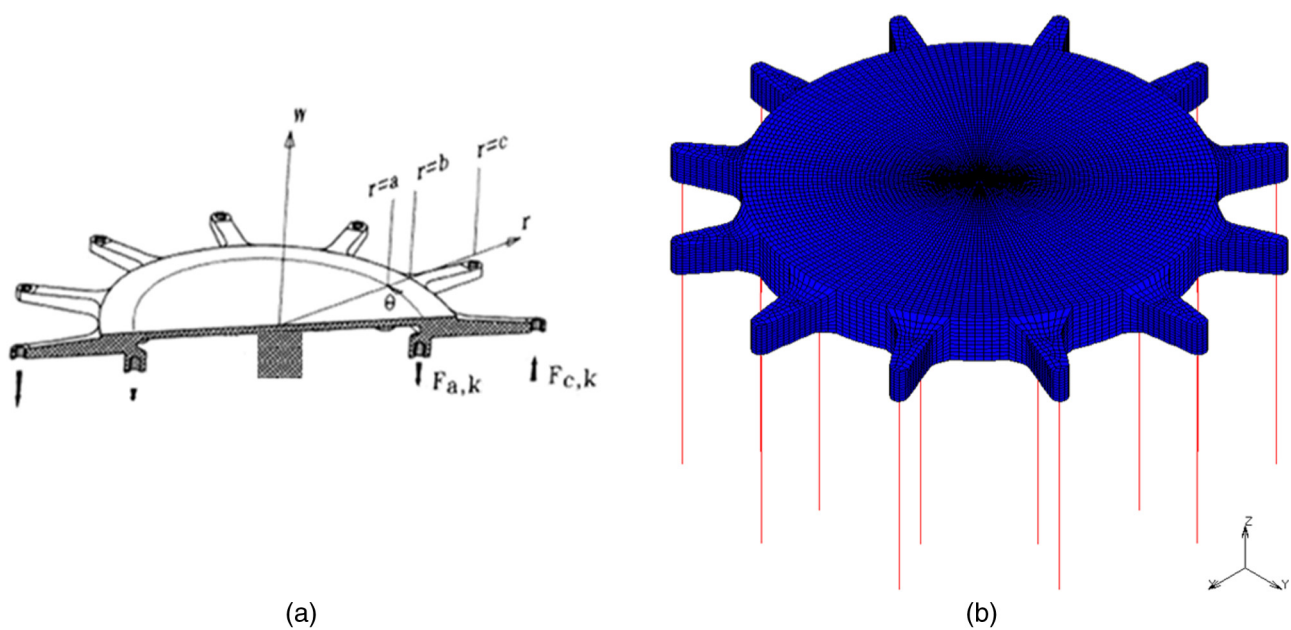

**Fig. 1** (a) Principle of a multimode deformable mirror (MDM). (b) Finite element model of the MADRAS mirror, with actuators represented by springs in red (63,708 hexaedral elements, 77,979 nodes –5881 nodes on the surface).

### 2.2.2 Optimization with finite element analysis

The mirror geometry is optimized with FEA for the generation of each specified mode. The parameters are the thickness of the central meniscus and the radii and thicknesses of the outer ring, the arms and the central clamping. For a given geometry, a finite element model is created and its 24 influence functions (IF) are recovered by applying a unit command to each actuator, while the others are fixed. Then, the correction of a given WFE $\phi_{in}$ is characterized by three main criteria:

1. The actuators' commands $\alpha$: given by the projection of the WFE on the set of influence functions' basis $B$:

$$\alpha = B^+ \phi_{in} = (B^t B)^{-1} B^t \phi_{in}. \quad (1)$$

2. The precision of correction: defined as the rms amplitude of the residual wavefront $\phi_{out}$ and determined by the reconstruction of the corrected wavefront $\phi_{cor}$:

$$\phi_{out} = \phi_{in} - \phi_{cor} = \phi_{in} - B(B^t B)^{-1} B^t \phi_{in}. \quad (2)$$

3. The resulting stress $\sigma$: determined by injecting the actuators commands into the finite element model.

The design optimization consists of minimizing these three criteria for each required mode. A classical least squares algorithm is used to converge to the optimal geometry.[31]

The finite element model, presented in Fig. 1, has 63,708 hexaedral elements and 77,979 nodes. The optical surface contains 100 nodes on a diameter and 120 angular sectors, providing sufficient sampling to characterize its deformation: up to 50 cycles per pupil. The actuators are modeled by springs with a given stiffness and length. The top extremity of the actuator is a node of the model, representing the force location (either under an arm or under the ring), and the bottom extremity is represented by a node clamped in the $(x, y)$ plane. A displacement along the $z$ axis is applied on the bottom spring node to simulate the actuation. The other boundary condition on the model is the central clamping: the bottom nodes of this part are fixed in the three directions.

### 2.3 FEA Characterization and Equipped System

The final mirror is 130 mm in diameter and 11 mm thick, with a 90 mm diameter, 3.5 mm thick optical pupil. The finite element model allows a full characterization of the system: the precision of correction and mechanical behavior are determined, such as hardware and integration requirements.

#### 2.3.1 Eigen modes

The system eigen modes are deduced from the set of influence functions by performing a singular value decomposition.[32] The eigen modes, shown in Fig. 2, constitute an orthogonal basis, representing all of the deformations that the system can achieve. As expected, the modes are similar to the Zernike polynomials.

#### 2.3.2 Correction performance

*Specified Zernike mode correction.* Each specified Zernike, deduced from the expected deformation maps of primary mirrors in space (Table 1), is decomposed over the set of influence functions' basis, in order to characterize mirror correction performance (see Sec. 2.2.2). The correction of tip, tilt and focus will be addressed by a five degrees of freedom mechanism on the secondary mirror: it is simulated by adding three virtual influence functions corresponding to these modes.

Figure 3 presents the study of the astigmatism3 correction and the performance of the correction of each mode is summarized in Fig. 4 and Table 1. All the modes are corrected with a precision better than 5 nm rms, except for spherical3 and pentafoil9 which are slightly worse. Fore these two modes, the residuals are due to the central clamping and to a symmetry mismatch between the system and the mode. In conclusion, the modal correction efficiency has been demonstrated with FEA.

*Global WFE correction.* With the corrective performance of each mode the overall system performance can be deduced. A representative WFE $\phi_{in}$ will be composed of a random combination of the specified modes $\phi_{mode,i}$.

$$\phi_{in} = \sum \lambda_i \phi_{mode,i}, \quad (3)$$

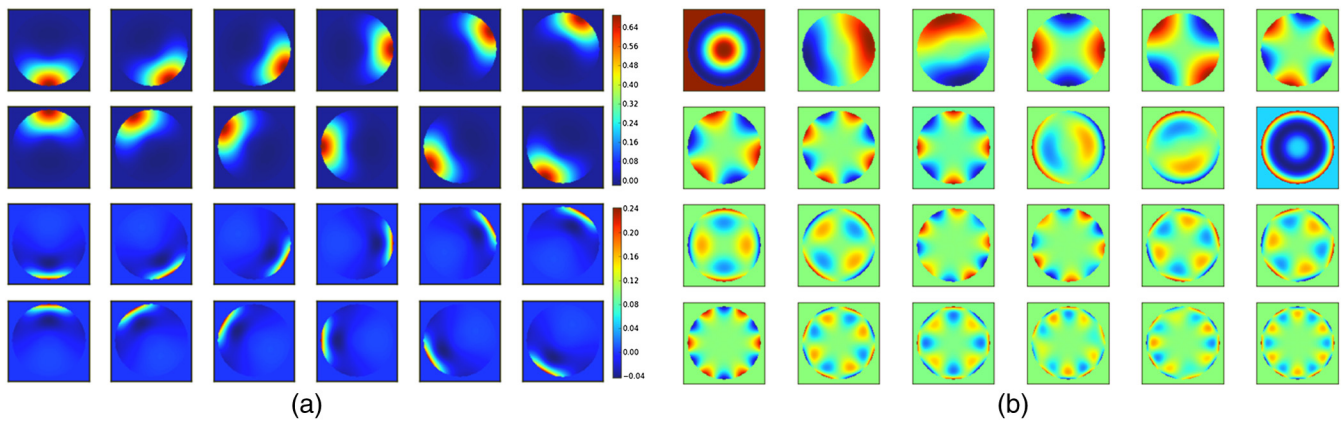

**Fig. 2** Finite element analysis (FEA) results: (a) system influence functions (unit: $\mu$m). (b) System eigen modes.

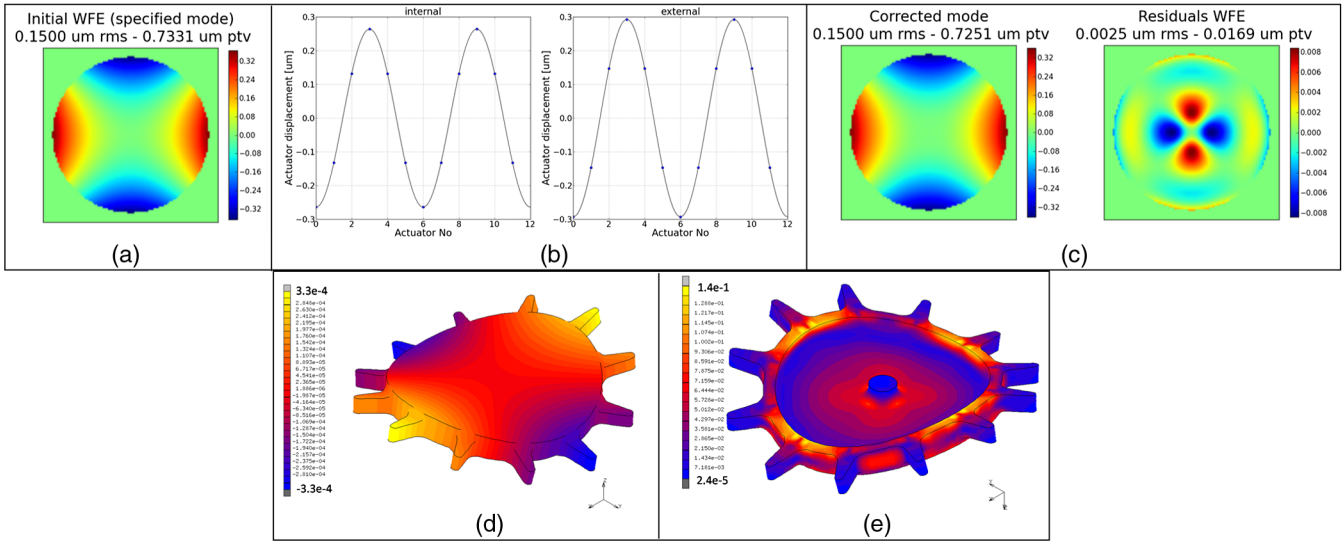

**Fig. 3** Characterization of the astigmatism3 correction: (a) Required mode (unit: $\mu$m). (b) Actuators' displacements, deduced from the mode projection on the influence functions. (c) Precision of correction: generated mode and residues (unit: $\mu$m). (d) Resulting mirror deformation, from FEA (unit: mm). (e) Resulting stress, from FEA (unit: MPa).

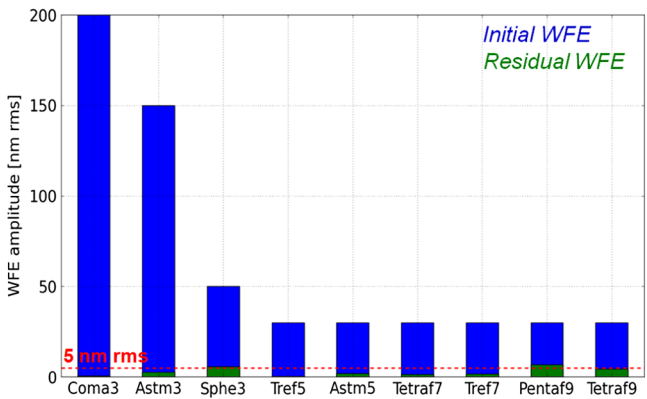

**Fig. 4** Correction of each specified mode: amplitude of the wavefront error before (in blue) and after (in green) correction.

with $\lambda$ a random coefficient between −1 and 1 and $\phi_{\text{mode}}$ the specified maximum amplitude (summarized in Table 1).

As the study comes within the context of linear mechanics, the corrective performance of each individual mode is simply added to determine the performance of a mode combination.

In the worst case, the system will have to correct the nine modes at their maximum specified amplitude, in the same orientation and at the same time. The study of this case gives the limiting performance of the system. The residual wavefront after the correction of the worst case is 10.7 nm rms, which is slightly above the 10 nm rms specified for the correction of a global wavefront. This result is acceptable due to its small likehood. The maximum level of stress is not located at the same place for each mode (on the arm, the actuator location or the center), so in the worst case, the maximum constraints will be lower than the sum of the values given for each mode, which is 5.10 MPa. The Zerodur elastic limit is considered to be around 10 MPa,[33] which gives a safety factor higher than two, ensuring the mechanical integrity of the system.

The global precision of correction is determined by performing a statistical study on the correction of 1000 random WFE, as defined in Eq. (3). The coefficient $\lambda_i$ gives the modes' amplitude; they are drawn from a uniform distribution. The expected mean precision is then 5.9 nm rms, with a standard deviation of 1.5 nm rms. The global correction efficiency is then within the 10 nm rms specification.

### 2.3.3 Reliability analysis: dead actuator impact

The characterization of the system's reliability can be performed by studying the impact of one or several dead actuators on the system performance. A dead actuator is not supplied any more, but keeps its stiffness. The occurrence of a dead actuator can be modeled in two different ways, depending if there is a system recalibration or not. Without recalibration, the mode to be corrected is still projected on the 24 influence functions but the command of the dead actuator is forced to 0. With a recalibration, the mode projection is carried out on the 23 remaining influence functions so the dead actuator will be compensated by its neighbors.

The impact of a dead actuator depends on the actuator location and on the mode to be corrected. The performance of correction is computed for each mode and for each actuator. The results advocate that a recalibration is essential to maintain reasonable performance. Without recalibration, the residual WFE is increased by a factor 7.7, on average; this factor is reduced to 1.3 with recalibration.

For a chosen azimuthal order $m$, modes with the greatest radial order $n$, are less influenced by the loss of one actuator. The corrections of the modes defined by $n = m$ are more damaged by the loss of an internal actuator, while the modes defined by $n = m + 2$ are more dependent on external actuators. This fact will allow balancing the impact of a dead actuator with respect to a global WFE correction. On the left of Fig. 5, the mean resulting residuals and the worst and best cases are compared to the nominal performance: with a recalibration, the loss of one actuator degrades the Laslandes et al.: Mirror actively deformed and regulated for applications in space...

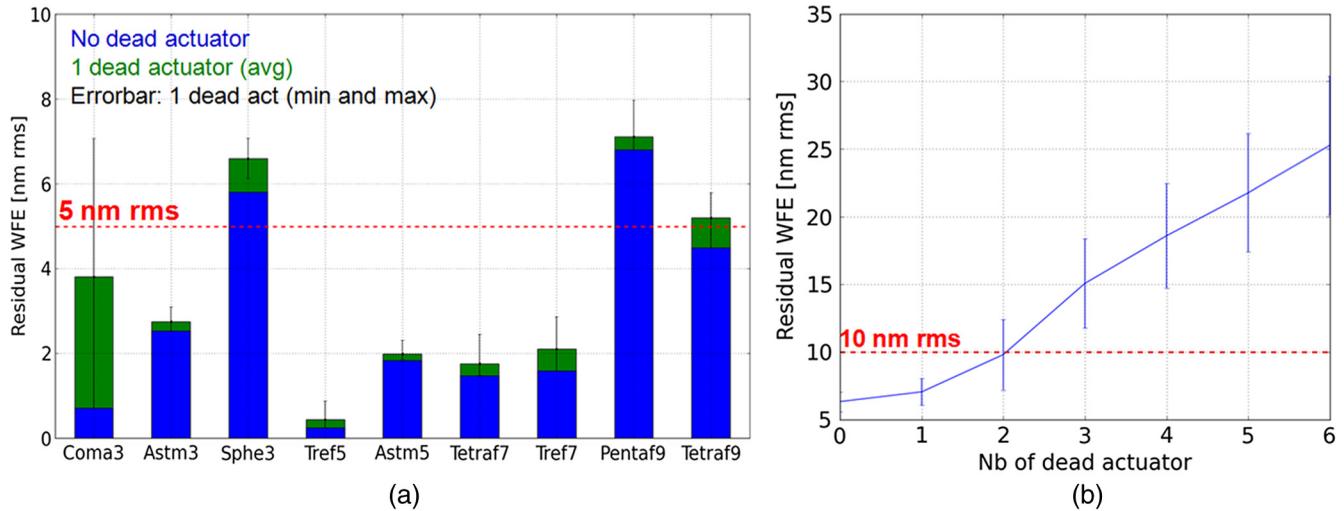

**Fig. 5** (a) System performance with one dead actuator (FEA results, considering a recalibration): comparison of the mean precision of correction of a system fully functional (in blue) and a system with one dead actuator (in green). (b) Evolution of the mean residual wavefront with the number of dead actuators, with a recalibration (statistics on 5000 random WFEs and sets of dead actuators).

performance in a reasonable way, the correction stays within the specification.

The evolution of the mean correction performance with the number of dead actuators can also be studied: for a given number of dead actuators, 50 random sets of dead actuators are drawn and the correction of 100 random WFEs is performed for each deteriorated set of influence functions basis (Fig. 5, right). Logically, the residues increase with the number of dead actuator, but we can see that with two dead actuators the system is still functioning within the specification: the mean precision is 9.7 nm rms, with a standard deviation of 2.1 nm rms.

This is possible due to the intrinsic redundancy of the system: 24 actuators are used for the generation of only 17 modes (the specified Zernike polynomials in any orientation). This performance is a major advantage for space applications, ensuring the robustness and reliability of the MADRAS concept.

### 2.3.4 Assembled system

*Definition.* The correcting system, shown in Fig. 6, weighs 4 kg and is 80 mm in height, with a diameter of 150 mm. It is composed of four main pieces:

1. The Zerodur mirror, with a central meniscus, an outer ring, 12 arms and a fixed central cylinder. It is polished in the 100 mm diameter aperture.

2. An Invar supporting structure, composed of a cone and a reference plate. The reference plate is stiffened with thin ribs so that any deformation of the plate will be negligible. The mirror is glued to the top part of the cone, on the bottom perimeter of the central clamp.

3. Twenty-four piezoelectric actuators. They are linked to the mirror and to the reference plate. Their required stroke is deduced from the worst case correction:

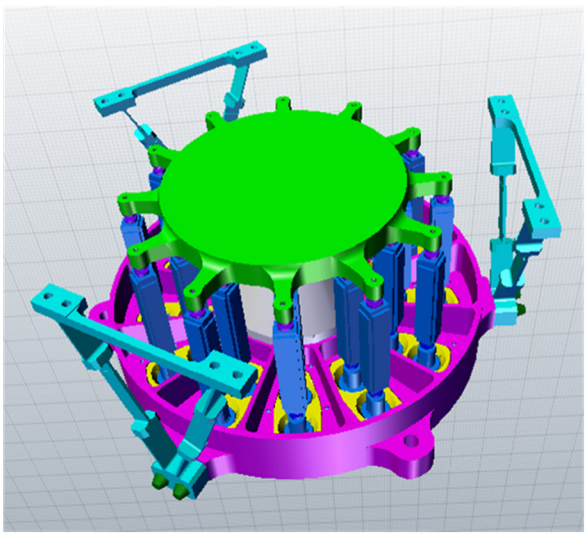
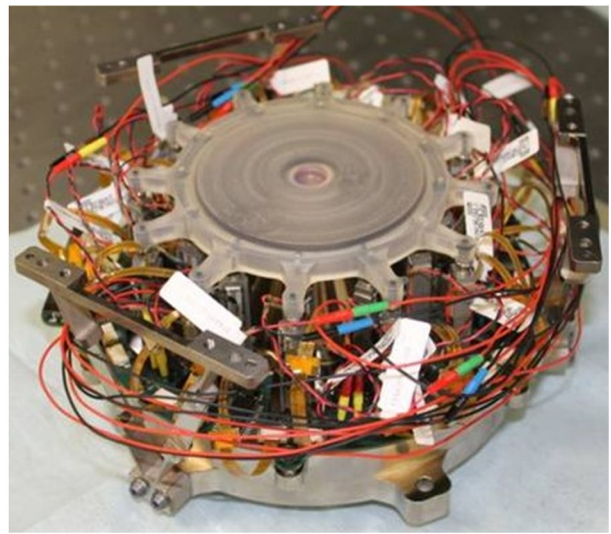

**Fig. 6** (a) Model of the assembled mirror (mirror in green, reference plate in purple, cone in gray, actuators in blue and fixation devices in light blue). (b) Integrated system.

- the actuators must provide a ±10 μm displacement in order to compensate for all the specified WFEs. The chosen actuators are the Cedrat PPA40M which have a 40 μm stroke, at 80 V.
4. The three bipods holding the system. These fixation devices, attached at three points on the reference plate, have been designed to provide an isostatic condition.

*Validation.* The assembled system is modeled with FEA, in order to verify that the structure does not change the mirror's performance or mechanical behavior. The main differences between the model of the assembled mirror and the model of the mirror alone come from the boundary conditions: the reference plate is not infinitely rigid but stiffened with ribs and fixed on three points, and the central pad is not completely clamped but glued on its periphery. The influence functions are recovered from the assembled mirror model and the study of each specified mode is performed, as in Sec. 2.3.2. This study has shown that the performance of the assembled mirror differs in a negligible way from the performance of the mirror alone. The characterization performed previously is still accurate and the assembled mirror design is validated.

Quasi-static and dynamic analyses are performed to simulate the system's mechanical behavior during launch. The mechanical strength of the system is estimated under quasi-static loads of 30 g, applied along three axes. For each direction, the maximum stress is 1 MPa, located around the mirror central clamp. During launch, this load is applied to the system in three directions, at the same time, so the maximum stress induced in the mirror will be 3 MPa, giving a safety factor of 3. A modal analysis has also been performed, giving the first mode at 233 Hz, which is a safe value considering the resonance frequencies of launch vehicles.[34] This validates the stiffness of the structure and the flexibility of the three fixation devices.

## 2.4 Conclusion of the Finite Element Analysis

Extensive FEA have allowed a complete opto-mechanical system definition, optimization and characterization. First, the mirror alone has been designed in order to correct the nine main Zernike modes expected to appear in a space telescope with a precision better than 5 nm rms. We also demonstrated that a representative WFE is corrected with a 6 nm rms precision. Secondly, the specifications and tolerancing on the system hardware have been defined and the assembled system has been designed and integrated.

## 3 Experimental Testing and Performance

### 3.1 Opto-Mechanical Validation

Once the system is integrated, a first performance characterization was conducted with a Fizeau interferometer in order to measure optical surface deformations. The mirror is facing downward and mounted on a tip/tilt plate. The goal of the measurement was to validate the mechanical design and the integration by correlating simulations and measurements. The interferometer has an optical aperture of 100 mm in diameter and a reference flat is used. In order to minimize the measurement noise, an averaging of nine measurements is systematically performed. The deformation maps are given on a grid of 550 by 550 pixels and the interferometer measurement precision is 1 nm.

The 24 system's influence functions are measured by sending a push/pull command to each actuator while the others are at rest. Their shape can be compared to the ones expected from FEA by normalizing the deformation maps. As we can see in Fig. 7, measured and simulated influence functions are very similar. As a result, the eigen modes have the expected Zernike shapes (see Fig. 8).

From the influence functions, the mirror correction capability can be deduced for each specified mode. With a difference lower than 1 nm between simulations and measurements, the expected precisions of correction are well recovered (see Fig. 9 and Table 1). In the worst case, the total residual error is 11.1 nm rms, and the mean precision, computed on 1000 random WFE, is 6.2 nm rms, with a standard deviation of 1.5 nm rms.

So the opto-mechanical design is validated, the mirror is able to efficiently generate the shapes that will allow the compensation of the expected deformation of a 3-m-class lightweight primary mirror in space. This interferometric characterization gives the ultimate performance of the system, which is the goal for the closed loop functioning.

### 3.2 Closed-Loop Performance

MADRAS mirror is then coupled to a wavefront Sensor (WFS) and a Real Time Computer (RTC) in order to demonstrate its functioning in closed loop in a representative configuration.

A telescope simulator injects a WFE onto the correcting mirror and the efficiency of the active loop is studied by comparing the measurements before and after correction. In order to fully characterize the system, several types of measurements are performed: wavefront sensing, point spread function (PSF) imaging and interferometry.

#### 3.2.1 Mirror control

For the demonstration, the wavefront sensing and control is performed through a Shack–Hartmann WFS. The sensor, placed in a plane conjugated to the active mirror, measures local wavefront slopes within subapertures defined by a microlenses array.[35] A Real-Time Computer processes the measurements and computes the mirror commands in order to match the measured wave-front with a reference one. Once the control law, loop gain and reference wavefront are defined, the active correcting loop operates autonomously, at a 1 Hz frequency.

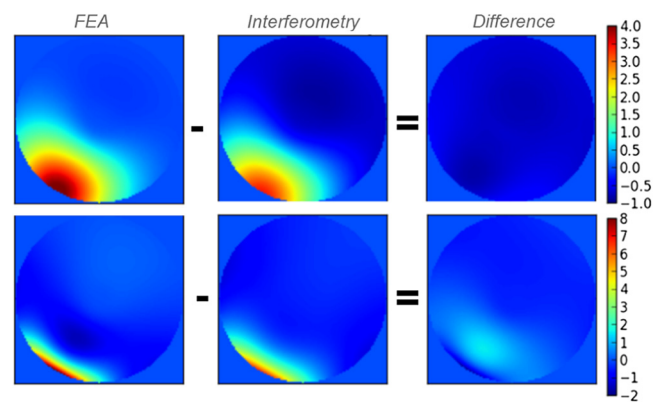

**Fig. 7** Comparison of the simulated and measured influence functions: top: internal actuator. Bottom: external actuator (both simulated and measured maps are normalized to a rms value of 1).

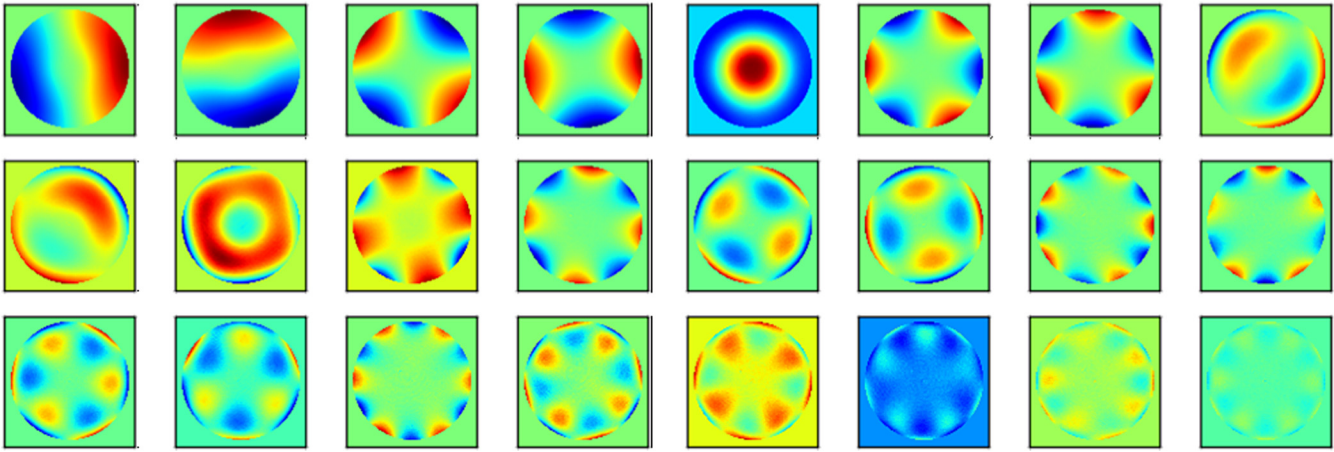

**Fig. 8** System eigen modes, deduced from the measured influence functions.

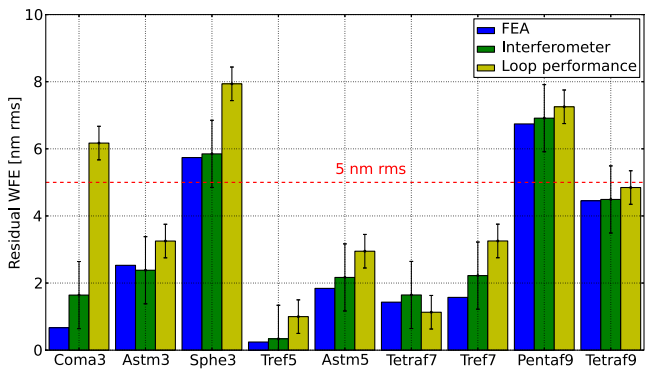

**Fig. 9** MADRAS performance: residual WFE measured by the wavefront sensor (WFS2) after the correction of each specified mode in closed loop (in pale green), compared to the expected performance, from interferometric measurements (in green) and from FEA (in blue) (errorbars correspond to the interferometer and WFS precisions).

This wavefront sensing and control strategy is typically used in adaptive optics systems.[36] But a different wavefront sensing approach could be adopted, optimized according to the mission.

### 3.2.2 Test bed

*Description.* The test bench, shown in Fig. 10, is composed of the following elements:

1. A point source and collimating lens, simulating the observation of a distant object.
2. A telescope simulator generating WFE expected in a 3-m-class telescope. It is an adaptive optics loop, composed of a 20-mm-diameter, 88-actuator magnetic DM (DM88[37]), a 3-mm-diameter Shack–Hartmann WFS1 with 784 subapertures and a RTC1. The DM pitch is 2.5 mm and the actuators are positioned on a 10 by 10 grid. This sampling allows the generation of all the required spatial frequencies and oversampling the WFS1 (28 by 28 subapertures) allows for an accurate measurement of the residual wavefront, up to 14 cycles per pupil.
3. An active correction loop, composed of the MADRAS mirror, a second 3 mm diameter $28 \times 28$ subaperture Shack–Hartmann WFS2 and a second RTC2.
4. Four beam expanders, relaying the pupil between the DM88, which is the entrance pupil, and the WFS1, the MADRAS mirror and the WFS2.
5. Two imaging cameras located in focal planes before and after the correction.
6. A Fizeau interferometer, directly looking at the mirror in order to monitor its deformation in real time.

*Calibration.* The active system calibration consists of performing an interaction matrix: the influence functions are measured with the WFS to compute a control matrix. As explained in Sec. 2.3.2, the external handling of tip, tilt and focus is simulated by adding virtual influence functions.

The loop noise is characterized by correcting the turbulent phase: a WFE is measured at $\pm 4.8$ nm rms. This precision is reduced to $\pm 0.5$ nm rms by averaging 100 measurements.

For an efficient PSF measurement, the first step is to correct the WFE seen by the WFS when a flat wavefront is injected. The contributors of this WFE have been identified and characterized:

1. The telescope simulator generates a flat wavefront and the specified modes with a precision of 5 nm rms.
2. Small test bench misalignment induces 18 nm rms of optical aberrations.
3. The integration bias induces a 200 nm rms WFE: at rest, the MADRAS mirror has some shape error, due to actuator integration.

The flattening is obtained with a residual error of 12.2 nm rms. In order to characterize the precision of correction of a calibrated WFE, this residual wavefront is the target for the next corrections.

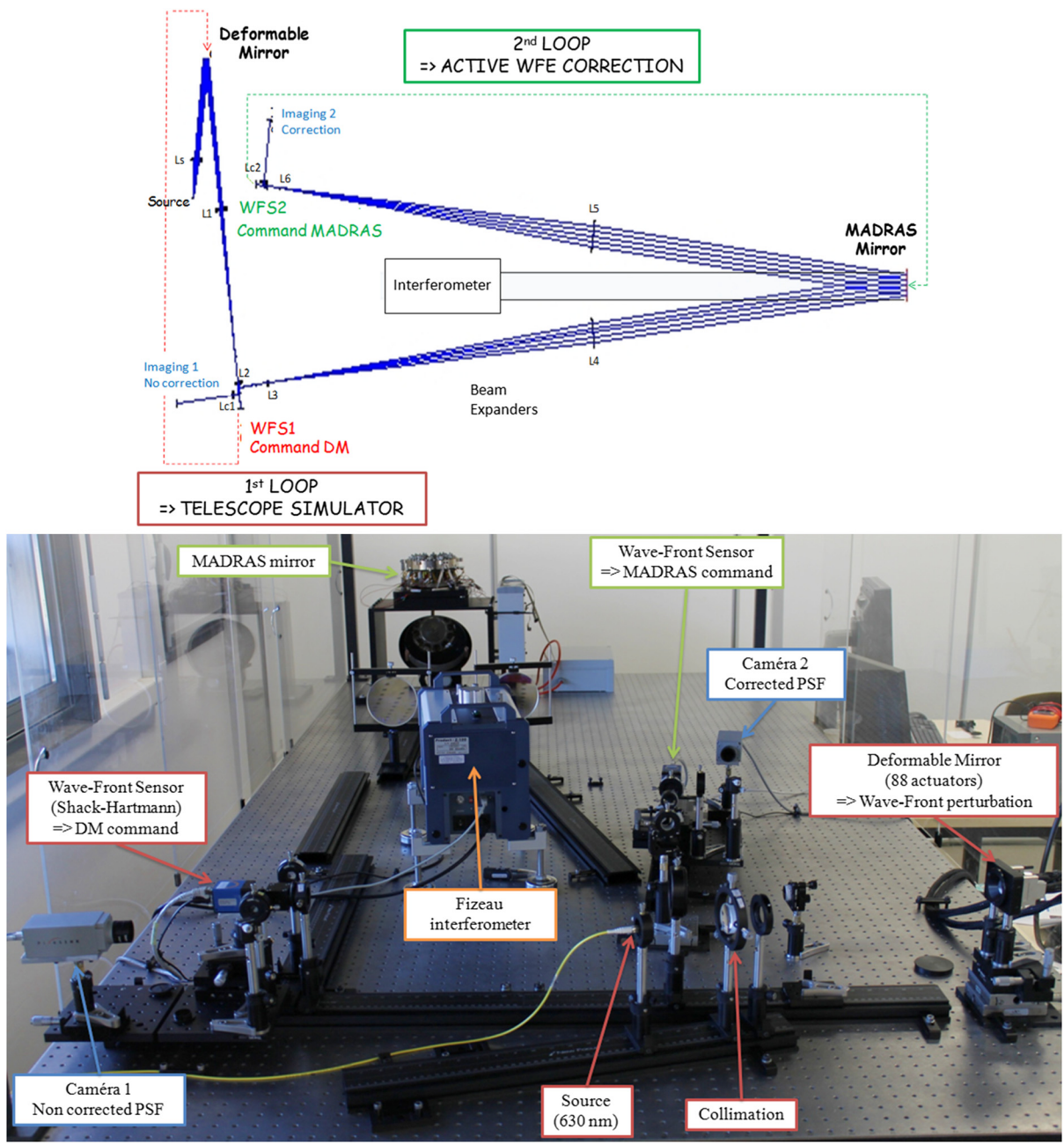

**Fig. 10** MADRAS testbed: optical design and picture.

### 3.2.3 Mode correction

The correction performance in closed loop is measured for each specified mode. They are all corrected with a precision better than 8 nm rms (see Fig. 9 and Table 1). The expected precision of correction, deduced from simulations and measurements with the Fizeau interferometer is recovered with a difference lower than 1.5 nm for seven of the modes.

The correction of astigmatism3&5, trefoil5&7 and tetrafoil7&9 is highly efficient, the residual WFE is below 5 nm rms, and the pentafoil9 correction precision is around 7 nm rms. The amplitude of the residuals measured for all these modes (except for tetrafoil7) is slightly higher than the expected one. This difference has several causes such as the numerical errors from the phase reconstruction using the measured slopes, and the noncommon path aberrations between the interferometric path and the MADRAS loop.

Coma3 and spherical3 are currently corrected with a precision of 6 and 8 nm rms, but this result could be improved up to the values expected from the interferometric measurements by working on the control matrix, or by adding a real system to actuate the tip, tilt and focus modes. Both coma and tilt modes have a component in $\cos(\theta)$ in their mathematical expressions, and both focus and spherical aberrations have a component in $\rho^2$. Thus, the generations of these modes are linked and the method used for the tip, tilt and focus handling impacts the correction. In a representative configuration, a five degrees of freedom mechanism will address the three modes. Tip, tilt and focus influence functions will then be real, automatically solving this problem. So, the mismatch between simulations and measurements for these two modes is not due to the active mirror itself but to the current control method. As seen in Sec. 3.1, the interferometric measurements have validated the efficient generation of these modes and, as for the other modes, the performance will be recovered with the new experimental set-up.

In Fig. 11 we present the wavefront shapes before and after the correction of the specified astigmatism3. The residual shape and amplitude are close to the ones expected from simulation (see Fig. 3). For each specified mode, the PSF before and after correction have been recorded, illustrating well the correction. The optical shape of the mirror has been measured with the Fizeau interferometer, constituting another validation of the correction: the measured deformation corresponds well to the injected WFE.

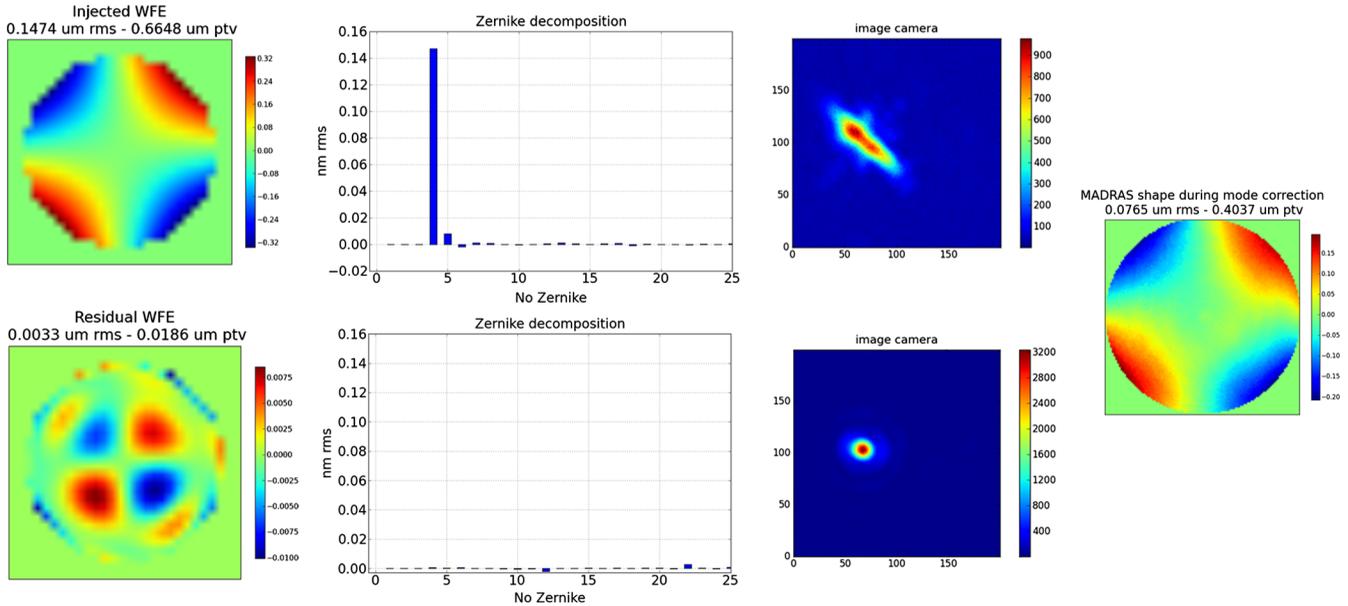

**Fig. 11** Astigmatism3 correction: WFE before (147.4 nm rms) and after (3.3 nm rms) correction—Zernike decomposition of the WFEs—point spread function (PSF) before and after correction—Optical shape of the mirror (76.5 nm rms).

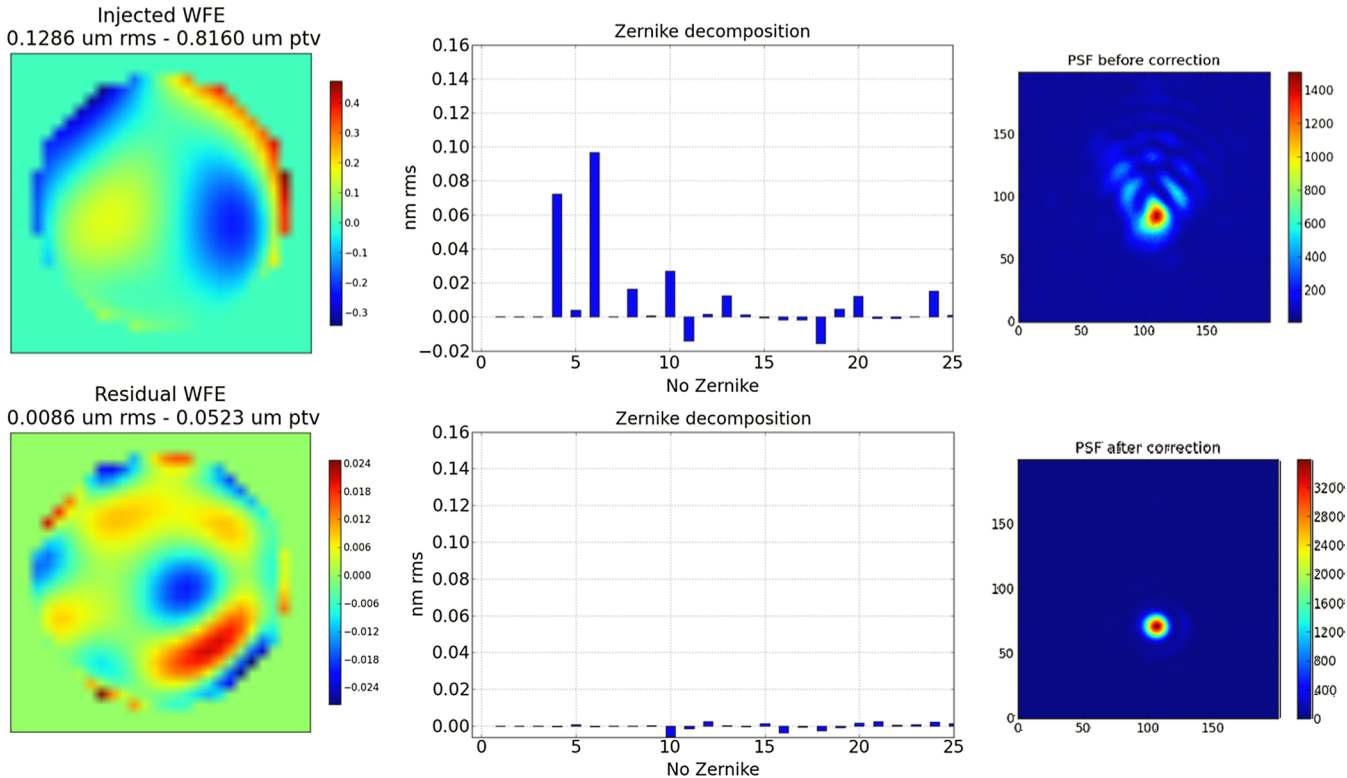

**Fig. 12** Correction of a random WFE (the given wavefront is an average of 100 measurements, tip, tilt and focus substracted).

In conclusion, the simulated precision of correction is efficiently recovered, which validates the functioning of the correcting mirror within the active loop.

### 3.2.4 Expected WFE correction

After having demonstrated the system's ability to correct each specified mode separately, the correction performance is studied regarding global WFE expected in space telescopes, as defined in Sec. 2.1. In order to perform a statistical study as in Sec. 2.3.2, the linearity of the system must be verified. This is validated by studying influence functions: each influence function amplitude evolves linearly with applied voltage, and the wavefront resulting from the command on several actuators corresponds to the sum of the wavefront resulting from individual commands. So, the experimental mean precision of the MADRAS system is deduced from a statistical study on the correction of 1000 random WFE, representative of a 3-m-class primary mirror deformation in space: 8.2 nm rms, with a standard deviation of 1.8 nm rms. This correction performance is well within the 10 nm rms specification.

By working on the tip, tilt and focus handling, this performance could be improved to 6.2 nm rms, deduced from the interferometric measurements. To reach this ultimate performance, a real tip, tilt and focus correction will be implemented on the test bench, with a motorized platform.

Finally, an example of random WFE correction is presented in Fig. 12: the injected wavefront is 129 nm rms and the corrected wavefront is $8.6 \pm 0.5$ nm rms. Once again, the gain for the PSF measurement is obvious.

## 4 Conclusion

The MADRAS project has demonstrated the performance of a correcting mirror dedicated to the compensation of large lightweight primary mirror shape errors in space. Instead of directly maintaining the primary mirror's optical shape, the correction is performed in the telescope exit pupil plane, allowing a reduction of the number of actuators and active system mass. The developed system has only 24 peripheral actuators, it is 130 mm diameter (for a pupil of 90 mm) and 80 mm high, it weighs 4 kg and has been conceived with regard to operation in space (low power consumption, low CTE, mechanical strength, robustness and reliability).

The first phase of the project consisted of a FEA optimization of the active mirror design, based on the correction requirements of a 3-m-class space telescope. The FEA has verified the mechanical strength of the system and helped define the specification on the hardware and the integration. The second phase consisted on the system assembly, integration and test. The opto-mechanical concept was validated with interferometric measurements: the measured influence functions are equivalent to the simulated ones along with the expected performance. Finally, the third phase allowed the complete characterization of the active system in a closed loop with relevant incoming perturbations. The correction performance has been experimentally demonstrated on a dedicated testbed: the developed system is able to compensate for specified deformations with a precision between 6 and 8 nm rms.

This project has brought the developed active mirror technology up to a Technology Readiness Level of 4, improving its maturity for space qualification. Its functioning has been fully validated in a laboratory environment. The next step is to perform environmental testing to reach a TRL5.

Placed in a pupil relay of future large telescopes, such a system will provide high resolution images while relaxing the tolerances on the assembly, integration and test phases. It will have a strong impact on the ratio performance over cost reduction, regarding telescopes development. This would be a major innovation which will allow the emergence of large, lightweight and compact space telescope concepts.

*Acknowledgments*

This study was performed with the support of a Ph.D. grant from CNES (Centre National d'Etudes Spatiales) and Thales Alenia Space, within a project of the Research and Industry Optical Cluster "PopSud/Optitec."

*References*

1. M. Postman et al., "Advanced Technology Large-Aperture Space Telescope: science drivers and technology developments," *Opt. Eng.* **51**(1), 011007 (2012).
2. O. Guyon, "Limits of adaptive optics for high-contrast imaging," *Astrophys. J.* **629**(1), 592–614 (2005).
3. L. E. Strigari, J. S. Bullock, and M. Kaplinghat, "Determining the nature of dark matter with astrometry," *Astrophys. J. Lett.* **657**(1), L1–L4 (2007).
4. C. Latry and J.-M. Delvit, "Staggered arrays for high resolution earth observing systems," *Proc. SPIE* **7452**, 74520O (2009).
5. E. Hecht, *Optics*, 2nd Ed., Addison-Wesley Publishing Company, Boston (1987).
6. D. Baiocchi and H. P. Stahl, "Enabling future space telescopes: mirror technology review and development roadmap," in astro2010: The Astronomy and Astrophysics Decadal Survey, Technology Development Paper No. 23 (2009).
7. L. Feinberg et al., "Space telescope design considerations," *Opt. Eng.* **51**(1), 011006 (2012).
8. C. J. Burrows et al., "The imaging performance of the Hubble Space Telescope," *Astrophys. J.* **369**, L21–L25 (1991).
9. G. L. Pilbratt et al., "Herschel Space Observatory. An ESA facility for far-infrared and submillimetre astronomy," *Astron. Astrophys.* **518**, 6 (2010).
10. E. Sein et al., "A Φ 3.5 m diameter Sic telescope for Herschel mission," *Proc. SPIE* **4850**, 608–618 (2003).
11. J. Gardner et al., "The James Webb Space Telescope," *Space Sci. Rev.* **123**(4), 485–606 (2006).
12. P. A. Lightsey et al., "James Webb Space Telescope: large deployable cryogenic telescope in space," *Opt. Eng.* **51**(1), 011003 (2012).
13. B. K. McComas, "Configurable adaptive optics for the correction of space-based optical systems," Ph.D. Thesis, Univ. of Colorado at Boulder (2002).
14. S. Kendrew, "Lightweight deformable mirrors for ground- and space-based imaging systems," Ph.D. Thesis, Univ. College London (2006).
15. R. H. Freeman and J. E. Pearson, "Deformable mirrors for all seasons and reasons," *Appl. Opt.* **21**(4), 580–588 (1982).
16. R. Davies and M. Kasper, "Adaptive optics for astronomy," *Ann. Rev. Astron. Astrophys.* **50**, 305–351 (2012).
17. R. N. Wilson et al., "Active optics: IV. Set-up and performance of the optics of the ESO new technology telescope (NTT) in the observatory," *J. Mod. Opt.* **38**(2), 219–243 (1991).
18. E.-D. Knohl, "VLT primary support system," *Proc. SPIE* **2199**, 271–283 (1994).
19. J. C. Dainty, A. V. Koryabin, and A. V. Kudryashov, "Low-order adaptive deformable mirror," *Appl. Opt.* **37**(21), 4663–4668 (1998).
20. M. Ferrari, "Development of a variable curvature mirror for the delay lines of the VLT interferometer," *Astron. Astrophys.* **128**, 221–227 (1998).
21. D. Redding et al., "Wavefront sensing and control for a Next-Generation Space Telescope," *Proc. SPIE* **3356**, 758–772 (1998).
22. D. Redding, D. Coulter, and J. Wellman, "Active optics for low-cost astronomical space telescopes," *Am. Astron. Soc. Meeting Abstr.* **219**, 136.05 (2012).
23. L. E. Cohan and D. W. Miller, "Integrated modeling for design of light-weight, active mirrors," *Opt. Eng.* **50**(6), 063003 (2011).
24. K. Patterson, N. Yamamoto, and S. Pellegrino, "Thin deformable mirrors for a reconfigurable space telescope," in *53rd AIAA Structures, Structural Dynamics, and Materials Conference*, Hawai (2012).

25. European Space Agency, "Technology readiness levels handbook for space applications," Technical Report (2008).
26. V. Costes, G. Cassar, and L. Escarrat, "Optical design of a compact telescope for the next generation earth observation system," in *International Conference on Space Optics*, Ajaccio (2012).
27. G. R. Lemaître, "Active optics: vase or meniscus multimode mirrors and degenerated monomode configurations," *Meccanica* **40**(3), 233–249 (2005).
28. R. J. Noll, "Zernike polynomials and atmospheric turbulence," *J. Opt. Soc. Am.* **66**(3), 207–211 (1976).
29. S. P. Timoshenko and S. Woinowsky-Krieger, *Theory of Plates and Shells, Engineering Mechanics Series*, McGraw-Hill International Editions, New York (1959).
30. A. Saint Venant, *Résumé des lecons de Navier sur l'application de la mécanique*, Dunod, Paris (1881).
31. J. Bonnans et al., *Numerical Optimization: Theoretical and Practical Aspects*, Springer, New York (2009).
32. G. H. Golub and C. F. van Loan, *Matrix Computations*, Johns Hopkins University Press, Baltimore (1996).
33. P. Hartmann et al., "ZERODUR glass ceramics for high stress applications," *Proc. SPIE* **7425**, 74250M (2009).
34. CNES, *Cours de Technologie Spatiale—Technique et Technologies des Vehicules Spatiaux*, Editions CILF, Paris (1998).
35. R. G. Lane and M. Tallon, "Wave-front reconstruction using a Shack-Hartmann sensor," *Appl. Opt.* **31**(32), 6902–6908 (1992).
36. R. Tyson and B. Frazier, *Field Guide to Adaptive Optics*, SPIE Field Guides, Bellingham (2004).
37. F. Rooms and J. Charton, "Deformable magnetic mirrors for adaptive optics," *Photoniques* **27**, 38–40 (2007).

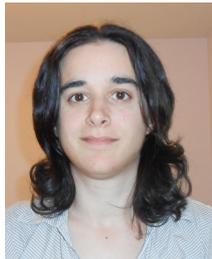

**Marie Laslandes** is a postdoctoral researcher specializing in astronomical optics at the California Institute of Technology in the Space Structure Laboratory. She is the recipient of a French ministry postdoctoral fellowship. Prior to this position, she was a PhD candidate at the Laboratoire d'Astrophysique de Marseille, funded by the French space agency. Her research interests include the design, characterization, and implementation of active systems for the next generation of space telescopes. She received a PhD in astronomical instrumentation from Aix-Marseille University in 2012.

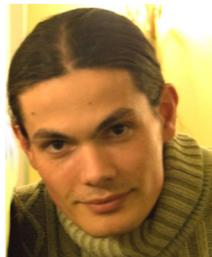

**Emmanuel Hugot** is a researcher at Laboratoire d'Astrophysique de Marseille. Its R&D activities are dedicated to Active Optics and innovative instrumentation. Involved in several projects for ground-based observatories for years, he has now oriented his work on the high angular resolution and high contrast imaging for space missions, specifically through the development of innovative focal plane architectures based on active mirrors and freeform optics. He received a PhD in astronomical instrumentation from Aix-Marseille University in 2007.

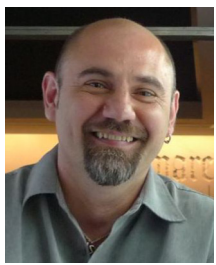

**Marc Ferrari** is an astronomer at Laboratoire d'Astrophysique de Marseille. He received a PhD in astronomical instrumentation from Aix-Marseille University in 1994. After positions at ESA and ESO, he joined LAM in 2000. A specialist on active/adaptive optics and optical fabrication, he was the optics R&D group leader from 2004 to 2011. Since 2012, he has been deputy director at LAM.

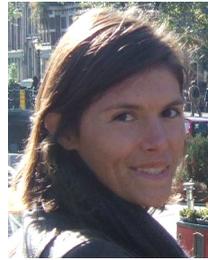

**Claire Hourtoule** is an engineer at Laboratoire d'Astrophysique de Marseille. She received her MS degree in astronomical instrumentation from Paris Observatory in 2004. She specializes in the control and testing of deformable mirror.

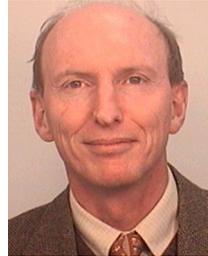

**Christian Singer** is the head of Mechanical, Thermal, and Optical Engineering Department at Thales Alenia Space (former Aerospatiale) since 2005. He graduated from Ecole Suprieure d'Optique in 1982. He joined Aerospatiale in 1985 and has worked in space science for more than 25 years. He was project manager on programs development such as ISO (Infrared Space Observatory for ESA), IASI (Infrared Atmospheric Sounding Interferometer instrument mounted on METOP) for the definition phase, Vegetation 2 (wide field of view imager mounted on SPOT5) and Planck (ESA) for the payload definition phase. Since 2000, he has been charge of advance studies.

**Christophe Devilliers** is an engineer at Thales Alenia Space in the Mechanical, Thermal, and Optical Engineering Department. He specializes in mechanical conception of space structures.

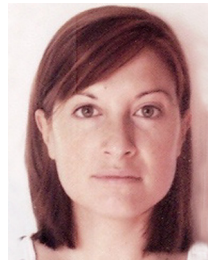

**Céline Lopez** is project manager at Thales SESO in Aix-en-Provence. She works on the design and manufacturing of optics for synchrotrons and satellites. She is currently working on components for the next generation of Meteosat. She graduated in 2006 from the National Engineering School of Tarbes in France and worked for different companies in aerospace; automotive; oil and gas; and nuclear domains.

**Frédéric Chazallet** is the creator and head of the Shakti company, specializing in electronics, calculators, and signal processing.